\documentclass[journal]{IEEEtran}

\IEEEoverridecommandlockouts

\usepackage[fleqn]{amsmath}
\usepackage{amssymb,graphicx,verbatim,enumitem,mdframed,mathrsfs}
\usepackage[numbers]{natbib}

\newtheorem{theorem}{\textbf{Theorem}}
\newtheorem{lemma}{\textbf{Lemma}}
\newtheorem{corollary}{\textbf{Corollary}}
\newtheorem{definition}{\textbf{Definition}}

\providecommand{\mnorm}[1]{|#1|}
\providecommand{\norm}[1]{\lVert#1\rVert}

\title{\LARGE \bf Privacy of Agents' Costs in Peer-to-Peer Distributed Optimization}

\author{Nirupam Gupta and Nikhil Chopra
\thanks{This work was supported by
the Naval Air Warfare Center Aircraft Division, Pax River, MD, under contract N00421132M022.}
\thanks{Nirupam Gupta ({\tt\small nirupam@umd.edu}) and Nikhil Chopra ({\tt\small nchopra@umd.edu}) are with the Department of Mechanical Engineering,
        University of Maryland, College Park, 20742 MD, USA}
}

\begin{document}

\maketitle

\begin{abstract}                
In this paper, we propose a protocol that preserves (statistical) privacy of agents' costs in peer-to-peer distributed optimization against a passive adversary that corrupts certain number of agents in the network. The proposed protocol guarantees privacy of the affine parts of the honest agents' costs (agents that are not corrupted by the adversary) if the corrupted agents do not form a vertex cut of the underlying communication topology. Therefore, if the (passive) adversary corrupts at most $t$ arbitrary agents in the network then the proposed protocol can preserve the privacy of the affine parts of the remaining honest agents' costs if the communication topology has $(t+1)$-connectivity. The proposed privacy protocol is a composition of a privacy mechanism (we propose) with any (non-private) distributed optimization algorithm. 

\end{abstract}
\begin{IEEEkeywords}
privacy; distributed optimization
\end{IEEEkeywords}

\section{Introduction}
\label{sec:intro}

A peer-to-peer distributed optimization algorithm refers to joint optimization of the aggregate of agents' costs in a peer-to-peer network~\cite{tsitsiklis1986distributed, nedic2001distributed, terelius2011decentralized}. 
It has been shown that in certain distributed optimization algorithms, an adversary (passive) can learn about all the agents' costs by corrupting a (non-trivial) subset of agents in the network~\cite{yan2013distributed, gade2016private, lou2017privacy}. This is clearly undesirable, especially in cases where agents' costs carry sensitive information, such as the economic dispatch problem in power grids~\cite{yang2013consensus} or distributed statistical learning over private data sets~\cite{boyd2011distributed}. 

 
In this paper, we propose a distributed optimization protocol that guarantees privacy of the \emph{affine} parts of the honest agents' (that are not corrupted by the adversary) costs against passive adversaries that corrupt agents that do not constitute a vertex cut of the underlying communication topology of the network. Expectedly, this sufficient condition is coherent with the results by Yan et al.~\cite{yan2013distributed}; the cost of an honest agent can be kept private against a set of passively adversarial agents (agents that are corrupted by a passive adversary) in the consensus-based subgradient distributed optimization protocol if and only if the honest agent has at least one honest agent as its neighbor in the underlying commnication topology. 

There exists differentially private (ref.~\cite{dwork2014algorithmic}) distributed optimization protocols~\cite{nozari2018differentially, huang2015differentially}. However, it should be noted that in order to ensure differential privacy agents must compute a random approximation of the optimal solution of the optimization problem~\cite{nozari2018differentially}. In this paper, we are interested in privacy protocols that only prevents loss of privacy (defined later) of honest agents' costs due to the exchange of information between agents during a distributed optimization protocol, and is not concerned with privacy loss due to the disclosure of the final optimal solution.

Existing secure multiparty computation based methods, for preserving privacy of honest agents' costs in distributed optimization algorithms, require the communication topology to be complete~\cite{catrina2010secure,li2006secure,toft2009solving}, else they rely on secure message transmission between agents to emulate the complete communication topology~\cite{garay2008almost}. The protocol proposed in this paper is effective even when the communication topology is not complete and only requires an agent to be aware of its neighboring agents (and not the entire network). Homomorphic encryption based privacy approaches~\cite{silaghi2004distributed, hong2016privacy} rely on computational intractability of known hard problems, such as the decisional composite residuosity problem or the RSA problem.  Thus, homomorphic encryption based techniques are effective if the computational power of the adversary is assumed bounded, which is true in practice, at least for now. In this paper, we are interested in statistical (unconditional) privacy that should hold regardless of the computational power of the passive adversary. Specifically, in the proposed protocol the adversary is unable to distinguish (quantified using KL-divergence~\cite{kullback1951information}) between any two possible set of costs of the honest agents that yield the same optimal solution. However, we note that homomorphic encryption based privacy methods can also protect privacy of agents' costs against an adversary that is eavesdropping on the communication links between the agents, unlike the protocol proposed in this paper.

Lou et al.~\cite{lou2017privacy} proposed a heterogeneous step-size consensus-based subgradient algorithm to preserve the privacy of the agents' costs. However,~\cite{lou2017privacy} has shown convergence (to an optimal solution of the optimization problem) of their algorithm only if the individual agents' costs are convex. It is known that individual agents' costs need not be convex as long as their aggregate is convex for a distributed optimization protocol to converge to the optimal solution~\cite{gade2016distributed}. As a consequence of this, the privacy protocol in this paper also does not rely on the convexity of agents' costs.

In a closely related work, Gade and Vaidya~\cite{gade2016private} have proposed a similar privacy approach wherein each agent adds \emph{correlated} random functions to their original costs and use these effective costs for solving the original optimization problem. The addition of correlated random functions preserves the global aggregate cost and the privacy of honest agents' costs is preserved as long as the adversarial agents that do not constitute a vertex cut in the communication topology. The privacy analysis in~\cite{gade2016private} uses the argument of \emph{compatibility} of honest agents' costs in \emph{view} of the adversarial agents. Specifically, it is argued that every set of honest agents' costs that is \emph{compatible} with the optimal solution could be the set of honest agents' costs as far as the adversarial agents (that do not form a vertex cut) are concerned. However, this privacy analysis does not consider the fact that probability distribution of random functions that are added to the original costs are also part of the adversary's information, as adversarial agents are also following the prescribed protocol, and creates a discrepancy between the priori (before the execution of privacy protocol) and posteriori (after the execution of the privacy protocol) probability distribution of honest agents' costs. In this paper, the probability distribution of the random functions that are added to the original costs is formally defined and assumed known to the adversarial agents. Additionally, we define a formal quantification of privacy (or loss of privacy) of agents' costs using the concept of indistinguishability based on KL-divergence~\cite{gupta2017privacy}. The defined privacy measure provides an additional insight on the dependence of privacy of agents' costs in peer-to-peer distributed optimization on the vertex expansion (ref. \cite{chung1996laplacians, marsden2013eigenvalues}) of the underlying communication topology.

Privacy techniques based on algebraic transformations have also been proposed recently, but only for privacy in multi-agent optimization where all the agents communicate with a central server to compute the solution of the optimization problem~\cite{hong2016privacy, mangasarian2011privacy, weeraddana2013per}. Most of these works are concerned about privacy in linear programming, however there do exist some work for privacy in a more general convex optimization problems~\cite{weeraddana2013per}. In this paper, we are interested in the peer-to-peer distributed optimization where agents only communicate with each other to compute the solution of the optimization problem without any central server.

\subsection{Summary of Our Contribution}

We present a general approach for achieving privacy in distributed optimization protocols with. Our approach constitutes of two phases:
\begin{enumerate}
    \item 
In the first phase, each agent shares correlated random values with its neighbors and then computes a new, `effective cost' based on its original cost and the random values shared with its neighbors. 

\item In the second phase, the agents run an arbitrary distributed optimization protocol to optimize the aggregate of their `effective costs', instead of their original costs, computed in the first phase.
\end{enumerate}
The first phase is designed to ensure that the aggregate of all agents' effective costs is identical to the aggregate of the agents' original costs. Therefore, the above two-phase approach does not affect the optimality of the obtained solution. Furthermore, the privacy holds in our approach---in a formal sense and under certain conditions, as discussed below---regardless of the distributed optimization protocol used in the second phase.
To prove this we consider the worst-case scenario where all the effective costs of the honest agents are revealed to all the agents (including the adversarial agents) in the second phase.

Speaking informally, the privacy guarantee is that the (colluding) adversarial agents learn very less (quantified later) about the collective affine parts of the original costs of honest agents from an execution of the protocol (described above) other than the aggregate of the affine parts of the honest agents' costs. This holds regardless of any prior knowledge the adversarial agents may have about the costs of (some of) the honest agents. 
We prove that our protocol satisfies this notion of privacy as long as the set of colluding passively adversarial agents is not a vertex cut of the communication topology.

The privacy mechanism in the first phase is similar in structure to the privacy mechanism proposed in Gupta et al.~\cite{gupta2017privacy} for privacy in distributed average consensus. 


\section{Notation and Preliminaries}
\label{sec:not}

\def\Z{{\mathbb Z}}
\def\R{{\mathbb R}}
\def\N{{\mathcal N}}

We let $\mathbb{R}$, $\R^n$ and $\R^{m \times n}$ denote the set real numbers, $n$-dimensional real-valued vectors and $m \times n$-dimensional real-valued matrices. $[m] = \{1,\ldots, \, m\}$. We use $1_n$ to denote the $n$-dimensional vector all of whose elements is~$1$. For $M \in \R^{n \times n}$, we denote its generalized inverse by $M^{\dagger}$ and its pseudo-determinant\footnote{Pseudo-determinant of a square real-valued matrix is equal to the product of its non-zero eigenvalues.} by $\text{det}^{*}(M)$. For any vector $x \in \R^n$, $Diag(x)$ represents a diagonal matrix of dimension $n \times n$ with the elements along the diagonal given by $x$.\\

\label{sub:gt}
We consider communication networks represented by  simple, undirected graphs. That is, the communication links in a network of $n$ agents is modeled via a 
graph 
$\mathcal{G} = \{\mathcal{V}, \,\mathcal{E} \}$ where
the nodes $\mathcal{V} \triangleq \{1,\ldots,n\}$ denote the agents, and there is an edge $\{i, j\} \in \mathcal{E}$ iff there is a direct communication channel between agents $i$ and~$j$.
We let $\N_i$ denote the set of neighbors of an agent $i \in \mathcal{V}$, i.e., $j \in \N_i$ if and only if $\{i,\,j\} \in \mathcal{E}$. (Note that $i \not \in \N_i$ since $\mathcal{G}$ is a simple graph.) 


We say two agents $i, j$ are \emph{connected} if there is a path from~$i$ to~$j$; since we consider undirected graphs, this notion is symmetric. A graph $\mathcal{G}$ is \emph{connected} if every distinct pair of nodes is connected; note that a single-node graph is connected.

\begin{definition}{(Vertex cut)}
A set of nodes $S \subset \mathcal{V}$ is a \emph{vertex cut} of a graph $\mathcal{G}=\{\mathcal{V}, \mathcal{E}\}$ if removing the nodes in~$S$ (and the edges incident to those nodes) renders the resulting graph unconnected. In this case, we say that $S$ \emph{cuts}~$\mathcal{V}\setminus S$.
\end{definition} 

A graph is \emph{$k$-connected} if the smallest vertex cut of the graph contains~$k$ nodes.

\def\G{\mathcal{G}}
\def\V{\mathcal{V}}
\def\E{\mathcal{E}}

Let $\G =\{\V, \E\}$ be a graph. The \emph{subgraph induced by $\V' \subset \V$} is the graph $\G' = \{\V', \E'\}$ where $\E' \subset \E$ is the set of edges entirely within $\V'$ (i.e., $\E' = \{\{i, j\} \in \E \mid i, j \in \V'\}$). We say 
a graph $\mathcal{G}=\{\mathcal{V}, \mathcal{E}\}$ has \emph{$c$ connected components} if its vertex set $\V$ can be partitioned into disjoint sets $\V_1, \ldots, \V_c$ such that (1)~$\G$ has no edges between $\V_i$ and $\V_j$ for $i \neq j$ and (2)~for all~$i$, the subgraph induced by $\V_i$ is connected.
Clearly, if $\mathcal{G}$ is connected then it has one connected component.

For a graph $\G=\{\V, \E\}$, we define its \emph{incidence matrix}
$\nabla \in \{-1,0,1\}^{\mnorm{\mathcal{V}}\times \mnorm{\mathcal{E}}}$ (see \cite{godsil2001algebraic})
to be the matrix with $|\V|$ rows and $|\E|$ columns in which
\[
	\nabla_{i,\,e} = \left\{\begin{array}{cl}1 & \hspace*{3pt} \text{if } e = \{i,\,j\} \text{ and } i < j\\ -1 &  \hspace*{3pt} \text{if } e = \{i,\,j\} \text{ and } i > j \\ 0 &  \hspace*{3pt} \text{otherwise.}\end{array}\right. 
\]
Note that $1_n^T\cdot \nabla = 0$. We use $\nabla_{*,e}$ to denote the column of $\nabla$ corresponding to the edge $e \in \mathcal{E}$.

We rely on the following result \cite[Theorem 8.3.1]{godsil2001algebraic}:
\begin{lemma}
\label{lem:o_m}
Let $\G$ be an $n$-node graph with incidence matrix $\nabla$. Then
$\text{rank}(\nabla) = n-c$, where $c$ is the number of connected components of~$\mathcal{G}$. 
\end{lemma}

\def\L{\mathcal{L}} 
The graph-Laplacian $\L$ is given as $\L = \nabla \nabla^T$. Let, $\mu_1 \geq \ldots, \geq \mu_{n-c} > 0$ denote the non-zero eigenvalues of $\L$. (Eigenvalues of $\L$ are real values because $\L$ is a symmetric real-valued matrix.) From Spectral theorem we can decompose the graph-Laplacian $\L$ as following:
\[\L = U Diag([\mu_1, \ldots, \, \mu_{n-c}, 0_c^T]^T)U^T\]
where, $U \in \R^{n \times n}$ is a unitary matrix constituting of the orthogonal eigenvectors of $\L$. For convenience, we denote the smallest non-zero eigenvalue of Laplacian $\L$ as $\underline{\mu}(\L)$

The generalized inverse $\L^{\dagger}$ of the graph-Laplacian of $\G$ with $c$ connected components is given as following(cf.~\cite{gutman2004generalized}):
\begin{align}
    \L^{\dagger} = U Diag([1/\mu_1, \ldots, \, 1/\mu_{n-c}, 0_c^T]^T) U^T \label{eqn:gen_inv}
\end{align}

\def\EE{\mathbb{E}}
\def\cov{\text{Cov}}
For a random variable (or vector) $r$, $f_{r}$ denotes its probability density function (or distribution). Let $\EE(r)$ denote its mean value. The covariance (or variance) of $r$ is denoted by $\cov(r)$, which is simply equal to $\EE(r^Tr)$. The relative entropy of two distributions $f_{r}$ and $f'_{r}$ is given by the KL-divergence:
\[D_{KL}(f_{r}, f'_{r}) = \int_{\mathcal{R}}f_{r}(s)\log\frac{f_{r}(s)}{f'_{r}(s)}ds\]
Here, $\mathcal{R}$ denotes the space of $r$.

\subsection{Problem Formulation}
\label{sec:prob_f}

We consider a network of $n$ agents where the communication network between agents is represented by an undirected, simple, connected graph~$\G=\{\V, \E\}$; that  is, agents $i$ and $j$ have a direct communication link between them iff $\{i, j\} \in \E$.
The communication link between two agents is assumed to be authenticated secure; equivalently, in our adversarial model we do not consider an adversary who can eavesdrop on communications between honest agents, or tamper with their communication. (Alternately, private and authentic communication can be ensured using existing standard cryptographic techniques.)

\def\X{\mathcal{X}}
Each agent $i$ holds a (private) cost~$h_i: \mathbb{R}^m \to \mathbb{R}$.  
A \emph{distributed optimization algorithm} is an interactive protocol that enables agents in the peer-to-peer network to cooperatively solve the following optimization problem
\begin{align}
    \underset{x \in \R^m}{\text{minimize}} \sum_i h_i(x) \label{eqn:opt}
\end{align}

\def\C{\mathcal{C}}
\def\H{\mathcal{H}}

We are interested in a distributed optimization protocol that ensures privacy of agents' costs against a passive adversary that corrupts some fraction of the agents in the network. 
We let $\C \subset \V$ denote the set of agents corrupted by the adversary, and let $\H = \V\setminus \C$ denote the remaining honest agents.
As stated earlier, we assume that the adversary is passive and thus runs the prescribed optimization protocol. Privacy requires that the entire \emph{view} of the adversary---i.e., the costs of the corrupted agents as well as their internal states and all the protocol messages they received throughout execution of the protocol---does not leak (significant) information about the original costs of the honest agents. Note that, by definition, the adversary learns $\sum_{i}h_{i}(x^*)$ (assuming it corrupts at least one agent) from which it can compute the sum of the costs of the honest agents at $x^*$, and so ideally we require that the adversary does not learn anything more about the honest agents' costs than this.

We relax the above privacy requirement and only focus on privacy of the affine parts of the agents' costs. This privacy requirement is formulated as follows. 

Every cost $h_i(x), \, i \in \V$ can be decomposed as 
\[h_i(x) = h_i^{(na)}(x) + h_i^{(a)}(x)\] 
where, $h_i^{(a)}(x)$ is affine in $x$ (referred as the affine part of $h_i$) and $h_i^{(na)}(x)$ is not affine in $x$. Specifically, 
\[h_i^{(a)}(x) = \alpha_{i}^T x, \, \forall x \in \R^m\]
where, $\alpha_{i} \in \mathbb{R}^m$. Note that we ignore the constants, as they do not make any contribution in the optimization problem \eqref{eqn:opt}. 

\def\dist{\text{ dist}}
Let $\alpha^k_i$ denote the $k$-th element of the affine coefficient of agent $i$ and $\alpha^k = [\alpha^k_1, \ldots, \alpha^k_n]^T$ be the vector of $k$-th elements of all the affine coefficients, for $k \in [m]$. Let $\alpha = \{\alpha_i\}_{i \in \V}$, then the distance between two sets of affine coefficients $\alpha, \, \alpha'$ is simply the aggregate Euclidean distance between their elements and is given by 
\[ \dist{}(\alpha, \, \alpha') = \sqrt{\sum_{k = 1}^m \norm{\alpha^k - \alpha'^k}^2}\]

Our privacy definition is formulated for the privacy of $\{\alpha_i\}_{i\in \H}$ (affine coefficients of honest agents' costs) against $\C$. Let $\alpha_\C = \{\alpha_i\}_{i \in \C}$ denote a set of affine coefficients of corrupted agents' costs, and $\alpha_\H = \{\alpha_i\}_{i \in \H}$ a set of affine coefficients of honest agents' costs.


\def\view{{\sf View}}

Fixing some protocol, we let $\view_\C(\alpha)$ be a random variable denoting the view of the corrupted agents in an execution of the protocol when all the agents begin with costs with affine coefficients~$\alpha$. (Specific form of $\view_\C(\alpha)$ is given later once the privacy protocol is described.)
Then:

\def\a{{(a)}}

\begin{definition}\label{def:ip}
A distributed optimization protocol is \emph{$(\C, \epsilon)$-affine private} if for all $\{\alpha_i\}, \{\alpha'_i\}$ such that  $\alpha_\C = \alpha'_\C$ and 
$\sum_{i \in \H} \alpha_i = \sum_{i \in \H} \alpha'_i$,
the relative entropy of the distributions of $\view_\C(\alpha)$ and $\view_\C(\alpha')$ is bounded above by $\epsilon > 0$ as
\[D_{KL}(\view_\C(\alpha), \,\view_\C(\alpha')) \leq \epsilon \dist(\alpha,\alpha')^2\]
\end{definition}
(Here, following the similar notation $\alpha' = [\alpha'_1,\ldots, \,\alpha'_m]$.)

We remark that this definition makes sense even if $|\C|=n-1$, though in that case the definition is vacuous since $\alpha_\H = \sum_{i \in \H} \alpha_i$ and so revealing the sum of the honest agents' costs reveals the affine coefficients of honest agent's cost!

In other words, the privacy definition above says that it is very difficult for the adversary to distinguish between two possible sets of the collective affine coefficients of honest agents, if the coefficients are close enough (closeness quantified by distance defined above) to each other and has the same joint aggregate.
\section{Private Distributed Optimization}
\label{sec:pm}

As described previously, our protocol has a two-phase structure. In the first phase, each agent~$i$ computes an ``effective cost'' $\widetilde h_i$ based on its original cost~$h_i$ and random affine costs it sends to its neighbors; this is done while ensuring that $\sum_i \widetilde h_i(x)$ is equal to $\sum_i h_i(x)$ for all $x \in \R^n$ (see below). In the second phase, the agents use any distributed optimization protocol $\Pi$ to solve the following optimization problem 
\begin{align}
    \underset{x \in \R^m}{\text{minimize}} \sum_i \widetilde h_i(x) \label{eqn:opt_new}
\end{align}
This (as will be shown) gives the solution to the original optimization problem \eqref{eqn:opt}. Also, in the first phase the effective cost of each agent is obtained by adding an affine cost to the original cost, thus $\widetilde h_i$ is convex if $h_i$ is convex. Hence, most of the existing distributed optimization (that assume individual costs to be convex) can be used to solve \eqref{eqn:opt_new}.

It may at first seem strange that we can prove privacy of our algorithm without knowing anything about the distributed optimization protocol $\Pi$ used in the second phase of our algorithm. We do this by making a ``worst-case'' assumption about $\Pi$, namely, that it simply reveals all the agents' costs to all the agents! Such an algorithm is, of course, not at all private; for our purposes, however, this does not immediately violate privacy because $\Pi$ is run on the agents' \emph{effective} costs~$\{\widetilde{h}_i\}$ rather than their true costs~$\{h_i\}$. 

From now on, then, we let the view of the adversary consist of the original costs of the corrupted agents, their internal states and all the protocol messages they receive throughout execution of the first phase of our protocol, and all the agents' effective costs~$\widetilde{h} = [\widetilde{h}_1, \ldots, \widetilde{h}_n]^T$ obtained at the end of the first phase. Our definition of privacy (cf.\ Definition~\ref{def:ip}) remains unchanged.

The first phase of our protocol proceeds as follows:
\begin{enumerate}
	\item Each agent $i \in \mathcal{V}$ chooses independent vectors $r_{ij} \sim N(0_m, \sigma^2 Diag(1_m))$ from $\R^m$ for all $j \in \mathcal{N}_i$, and sends $r_{ij}$ to agent~$j$. Here, $\sigma \in \R$.
	\item Each agent $i \in \mathcal{V}$ computes a mask
		\begin{align}
			a_i = \sum_{j \in \N_i}(r_{ji}-r_{ij}) , \label{eqn:i_vn}
		\end{align}
		where $a_i \in \R^m$.
	\item Each agent $i \in \mathcal{V}$ computes effective cost $\widetilde h_i$ such that
	    \begin{align}
		    \widetilde{h}_i (x) = h_i (x) + a_i^T x, \, \forall x \in \R^m. \label{eqn:eff_cost}
	    \end{align}
	    
\end{enumerate}

Note that
\[
\sum_i \widetilde h_i(x)  =  \sum_i h_i(x) + \sum_i a_i^T x 
\]
Moreover,
\begin{align*}
\sum_i a_i = \sum_i \sum_{j \in \N_i}(r_{ji}-r_{ij}) = 0
\end{align*}
since $\G$ is undirected. Thus, $\sum_i \widetilde h_i \equiv \sum_i h_i$ and hence correctness of our overall algorithm (i.e., including the second phase) follows. Also, as mentioned earlier the new effective costs are still convex if the original costs are convex.

Now, the effective affine coefficients of agent are given by $\{\widetilde{\alpha}_i\}$, where $\widetilde{\alpha}_i= \alpha_i + a_i, \, \forall i \in \V$. As the view of the adversary consists of all the effective costs $\{\widetilde{h}_i\}$, therefore from the above privacy protocol we can infer that the privacy of $\{h_i\}_{i \in \H}$ depends on the privacy of $\{\alpha_i\}$ given the values of $\{\widetilde{\alpha}_i\}$.

Our algorithm is illustrated by example in Section~\ref{sec:illus}. 



\subsection{Privacy Analysis}
\label{sec:pa}

We show here that $(\C,\epsilon)$-affine privacy holds as long as $\C$ is not a vertex cut of~$\G$.

\label{sub:masks}

For an edge $e=\{i,j\}$ in the graph with $i<j$,
define \[b_e = r_{ji}-r_{ij}.\]
Let $b^k =[b^k_{e_1}, \ldots]^T$ be the vector of the $k$-th elements of all such vectors $b_e$ for all the edges in~$\G$.
If we let $a^k=[a^k_1, \ldots, a^k_n]^T$ denote vector consisting of $k$-th elements of the masks used by the agents, then we have
\[a^k = \nabla \cdot b^k .\]
Since the vectors $r_{ij}$ are identical and independent with normal distribution $N(0_m, \Sigma)$, it is easy to see that the values $\{b^k_e\}_{e \in \E}$ are independent and have identical normal distribution $N(0,\sqrt{2}\sigma)$ in $\R$.
Thus, $a^k$ is normally distributed over $\R^n$ with mean value and covariance matrix equal to $0_n$ and $2\sigma^2 \L$, respectively for all $k \in [m]$. Specifically, the probability density of $a^k$ at any point $a \in \R^n$ is given as 
\begin{align}
	f_{a^k}(a) = \frac{1}{\sqrt{\det^{*}(4\pi\sigma^2\L)}}\exp\left(-\frac{1}{4\sigma^2}a^T\L^{\dagger}a\right) 
\end{align}
where, $\det^{*}(4\pi\sigma^2\L) = (4\pi\sigma^2)^{n-c}\prod_{i = 1}^{n-c}\mu_i$ (product of non-zero eigenvalues).
As $\text{rank}(\L) = n-1$ when $\G$ is connected, we have:
\begin{lemma}
\label{lem:dist_a}
If $\G$ is connected then $a^k$ is normally distributed over all points in $\R^{n}$ subject to the constraint that $\sum_i a^k_i = 0$ for all $k \in [m]$, with mean value $\EE(a^k) = 0_n$ and covariance $\cov(a^k) = 2\sigma^2\L$.
\end{lemma}

Since $\widetilde{\alpha}_i = \alpha_i + a_i$, we have. 
\begin{lemma}
\label{lem:cond_mask}
If $\G$ is connected then the $k$-th elements of the effective affine coefficients $\widetilde \alpha^k = [\widetilde \alpha^k_1, \ldots, \, \widetilde \alpha^k_n]^T$ are normally distributed in $\R^n$ subject to the constraint that $\sum_i \widetilde \alpha^k_i = \sum_i \alpha^k_i$ for all $k \in [m]$, with mean value $\EE(\widetilde \alpha^k) = \alpha^k =[\alpha^k_1, \ldots, \alpha^k_n]^T$ and covariance $\cov(\widetilde \alpha^k)  = 2\sigma^2\L$.
\end{lemma}
\begin{IEEEproof}
As $\widetilde{\alpha}_i = \alpha_i + a_i, \forall i \in \V$. Thus, $\widetilde \alpha^k_i = \alpha^k_i + a^k_i$ for every $i \in \V$ and $k \in [m]$. ($\widetilde \alpha^k_i$ and $\alpha^k_i$ denote the $k$-th elements of $\widetilde \alpha_i$ and $\alpha_i$, respectively.)

If we let $\widetilde \alpha^k = [\widetilde \alpha^k_1, \ldots, \widetilde \alpha^k_n]^T$ and $\alpha^k = [\alpha^k_1, \ldots, \alpha^k_n]^T$, then we can have $\widetilde \alpha^k = \alpha^k + a^k$. As $a^k$ is independent of $\alpha^k$, then for connected $\G$ from Lemma \ref{lem:dist_a} we conclude that $\sum_i \widetilde \alpha^k_i = \sum_i \alpha^k_i$, and $\widetilde{\alpha^k}$ is normally distributed under this constraint with $\EE(\widetilde \alpha^k) = \alpha^k$ and $\cov(\widetilde \alpha^k) = 2\sigma^2\L$.
\end{IEEEproof}

Let $f_{\widetilde{\alpha}|\alpha}$ denote the probability density function (or distribution) of the collective effective affine coefficients $\widetilde{\alpha} = [\widetilde{\alpha}_1, \ldots, \, \widetilde{\alpha}_n]$ given that the true affine coefficients are $\alpha = [\alpha_1, \ldots, \alpha_n]$. Then, using Lemma \ref{lem:cond_mask}, we get the following:

\begin{theorem}
\label{thm:priv_1}
If $\mathcal{G}$ is connected then 
\begin{align}
    D_{KL}(f_{\widetilde{\alpha}|\alpha}||f_{\widetilde{\alpha}|\alpha'}) \leq \epsilon \dist(\alpha,\alpha')^2
\end{align}
for any two sets of collective affine coefficients $\alpha = [\alpha_1, \ldots, \alpha_n]$ and $\alpha' = [\alpha'_1, \ldots, \alpha'_n]$ that satisfy the constraint $\sum_i \alpha_i = \sum_i \alpha'_i$. Here, $\epsilon = 1/(4\sigma^2\underline{\mu}(\L))$.
\end{theorem}
\begin{IEEEproof}
Throughout the proof, we assume that $\G$ is connected. 

As $\widetilde \alpha^k = \alpha^k + a^k$ and the value of $a^k$ is independent of $\alpha^k$ for every $k \in [m]$, thus
\begin{align*}
    f_{\widetilde \alpha^k | \alpha}(\widetilde \alpha^k) = f_{a^k}(\widetilde{\alpha}^k-\alpha^k)
\end{align*}
This implies (cf. Lemma \ref{lem:dist_a}),
\begin{align*}
    &\log \frac{f_{\widetilde \alpha^k | \alpha}(\widetilde \alpha^k)}{f_{\widetilde \alpha^k | \alpha'}(\widetilde \alpha^k)} = \frac{1}{4\sigma^2} \times \\
    & \left\{(\widetilde{\alpha}^k-\alpha^k)^T \L^{\dagger}(\alpha^k-\alpha^k) -  (\widetilde{\alpha}^k - \alpha'^k)^T \L^{\dagger} (\widetilde{\alpha}^k - \alpha'^k)\right\} 
\end{align*}
By further simplifying the above, we get
\begin{align*}
    \log \frac{f_{\widetilde \alpha^k | \alpha}(\widetilde \alpha^k)}{f_{\widetilde \alpha^k | \alpha'}(\widetilde \alpha^k)} = \frac{1}{4 \sigma^2} (\alpha^k - \alpha'^k)^T \L^{\dagger} (2\widetilde{\alpha}^k - \alpha^k - \alpha'^k)
\end{align*}
For simplicity, we let $a = \widetilde{\alpha}^k-\alpha^k$. Then,
\begin{align*}
    & D_{KL}(f_{\widetilde \alpha^k | \alpha} || f_{\widetilde \alpha^k | \alpha'}) = \\
    & \frac{1}{4\sigma^2}\int_{a \in \R^n}(\alpha - \alpha')^T \L^{\dagger} (2a + \alpha - \alpha')f_{a^k}(a)da = \\
    & \frac{1}{2\sigma^2}(\alpha^k - \alpha'^k)^T\L^{\dagger}\EE(a^k) + \frac{1}{4\sigma^2}(\alpha^k -  \alpha'^k)^T \L^{\dagger} (\alpha^k - \alpha'^k)\\
    & = \frac{1}{4\sigma^2}(\alpha^k - \alpha'^k)^T \L^{\dagger} (\alpha^k - \alpha'^k)  
\end{align*}
As $1^T_n(\alpha^k - \alpha'^k) = 0_n$, i.e. $\alpha^k - \alpha'^k$ is orthogonal to $1_n$ and $\text{rank}(\L) = n-1$, this implies
\begin{align*}
    D_{KL}(f_{\widetilde \alpha^k | \alpha} || f_{\widetilde \alpha^k | \alpha'}) \leq \frac{\norm{\alpha^k - \alpha'^k}^2}{4\sigma^2\mu_{n-1}} = \frac{\norm{\alpha^k - \alpha'^k}^2}{4\sigma^2\underline{\mu}(\L)}
\end{align*}
Note that different elements $a^k$ and $a^{k'}$ of the masks are independent of each other, where $k \neq k' \in [m]$. Therefore,
\begin{align*}
    f_{\widetilde \alpha | \alpha} = \prod_{k = 1}^m f_{\widetilde \alpha^k | \alpha} \text{ and similarly, }f_{\widetilde \alpha | \alpha'} = \prod_{k = 1}^m f_{\widetilde \alpha^k | \alpha'}
\end{align*}
Hence, 
\begin{align*}
    D_{KL}(f_{\widetilde \alpha | \alpha} || f_{\widetilde \alpha | \alpha'}) = \sum_{k = 1}^mD_{KL}(f_{\widetilde \alpha^k | \alpha} || f_{\widetilde \alpha^k | \alpha'}) \leq \frac{\dist(\alpha, \alpha')^2}{4\sigma^2\underline{\mu}(\L)}
\end{align*}
\end{IEEEproof}

The above implies $(\C, \epsilon)$ - affine privacy of the proposed distributed optimization algorithm for the case when $\C=\emptyset$, i.e., when there are no corrupted agents. In that case, the view of the adversary consists only of the effective affine coefficients~$\widetilde{\alpha}$, and Lemma~\ref{lem:cond_mask} shows that the distribution of those values depends only on the sum of the agents' true affine coefficients. 
Below, we extend this line of argument to the case of nonempty~$\C$.




\label{sub:suff}

Fix some set $\C$ of corrupted agents, and recall that $\H=\V\setminus \C$. Let $\E_\C$ denote the set of edges incident to~$\C$, and let $\E_{\H} = \E \setminus \E_\C$ be the edges incident only to honest agents. We refer to $\G_\H = \{\H, \E_\H\}$ as the \emph{honest graph} and let $\L_{H}$ denote the graph-Laplacian of $\G_\H$. 
Note that now the adversary's view contains (information that allows it to compute) $\{b_e\}_{e \in \E_\C}$ in addition to the honest agents' affine coefficients~$\{\widetilde \alpha_i\}_{i \in \H}$.

The key observation enabling a proof of privacy is that the values $\{b^k_e\}_{e \in \E_{\H}}$ are independent in $\R^{|\H|}$ \emph{even conditioned on the values of~$\{b_e\}_{e \in \E_\C}$}, for every $k \in [m]$. Thus, owing to Theorem \ref{thm:priv_1}, we get the following privacy guarantee:

\begin{theorem}
\label{thm:priv_2}
If $\C$ is not a vertex cut of $\G$, then our proposed distributed optimization protocol is $(\C, \epsilon)$-affine private, with $\epsilon = 1/(4\sigma^2 \underline{\mu}(\L_{\H}))$. 
\end{theorem}
\begin{IEEEproof}
Throughout the proof assume that $\G_\H$ is connected, as $\C$ does not cute $\H$.

For given $\alpha = \{\alpha_i\}_{i \in \V}$, 
\[\view_{\C}(\alpha) = \{\{\widetilde \alpha_i\}_{i \in \H}, \, \{b_e\}_{e \in \E_C} \} \]
Let, $\widetilde \alpha_{\H} = \{\widetilde \alpha_i\}_{i \in \H}$ and $b_{\E_C} = \{b_e\}_{e \in \E_C}$.

Consider two sets of affine coefficients $\alpha = \{\alpha_i\}_{i \in \V}$ and $\alpha' = \{\alpha'_i\}_{i \in \V}$ such that $\alpha_i = \alpha'_i, \, \forall i \in \C$ and $\sum_{i}\alpha_i = \sum_i \alpha'_i$.

Then from Theorem \ref{thm:priv_1}, we get
\begin{align*}
    D_{KL}(f_{\widetilde \alpha_{\H} | \alpha} || f_{\widetilde \alpha_{\H} | \alpha'}) \leq \epsilon \dist(\alpha, \alpha')^2
\end{align*}
where, $\epsilon = 1/(4\sigma^2 \underline{\mu}\L_{\H}))$ and $\underline{\mu}(\L_{\H})$ is the smallest non-zero eigenvalue of $\L_{\H}$.

As $\{b_e\}_{e \in \E_C}$ are chosen independently of the affine coefficients $\alpha$, the above implies,
\begin{align*}
    D_{KL}(\view_{\C}(\alpha) || \view_{\C}(\alpha')) \leq \epsilon \dist(\alpha, \alpha')^2
\end{align*}
\end{IEEEproof}

As a corollary, we have

\begin{corollary}
\label{cor:f}
If $\G$ is $(t+1)$-connected, then for any $\C$ with $|\C| \leq t$ our proposed distributed optimization protocol is $(\C,\epsilon)$-affine private.
\end{corollary}

The value of $\epsilon$ is a quantitative measure of privacy, and smaller is the value of $\epsilon$ higher is the privacy. As is shown in Theorem~\ref{thm:priv_2}, $\epsilon$ is inversely proportional to the variance of random values added to the affine coefficients, which is quite intuitive. As the proposed privacy mechanism does not affect the accuracy of the distributed optimization protocol, thus we can choose value of $\sigma$ appropriately for desirable privacy measure.

Also, it is interesting to note that the value of $\epsilon$ is inversely proportional to the smallest non-zero eigenvalue of the Laplacian of the graph $\L_\H$. From Cheeger's Theorem~\cite{cheeger1969lower}, we know that
\begin{align}
    \frac{\phi^2_{\G_\H}}{2} \leq \underline{\mu}(\L_{\H}) \leq 2 \phi_{\G_\H} \label{eqn:chee}
\end{align}
where, $\phi_{\G_\H}$ is a non-negative real number that is known as the \emph{vertex expansion} of graph $\G_\H$. Value of $\phi_{\G_\H}$ roughly indicates how close $\G_\H$ is to being not connected (specific form is omitted here, interested reader can refer to~\cite{chung1996laplacians, marsden2013eigenvalues}). In short, smaller is the value of $\phi_{\G_\H}$, lower is the vertex connectivity of $\G_\H$ and vice-versa. The value of $\phi_{\G_\H}$ is zero if and only if $\G_\H$ is not connected. Thus, owing to the Cheeger's inequality~\eqref{eqn:chee} the value of $\epsilon$ in Theorem~\ref{thm:priv_2} is bounded above as
\begin{align*}
    \epsilon \leq \frac{1}{2\sigma^2 \phi^2_{\G_\H}}
\end{align*}
This means that higher vertex expansion of the honest graph $\G_\H$ implies better privacy.  

\section{Illustration}
\label{sec:illus}
\begin{figure}[htb!]
\begin{center}
	\includegraphics[width=0.3\textwidth]{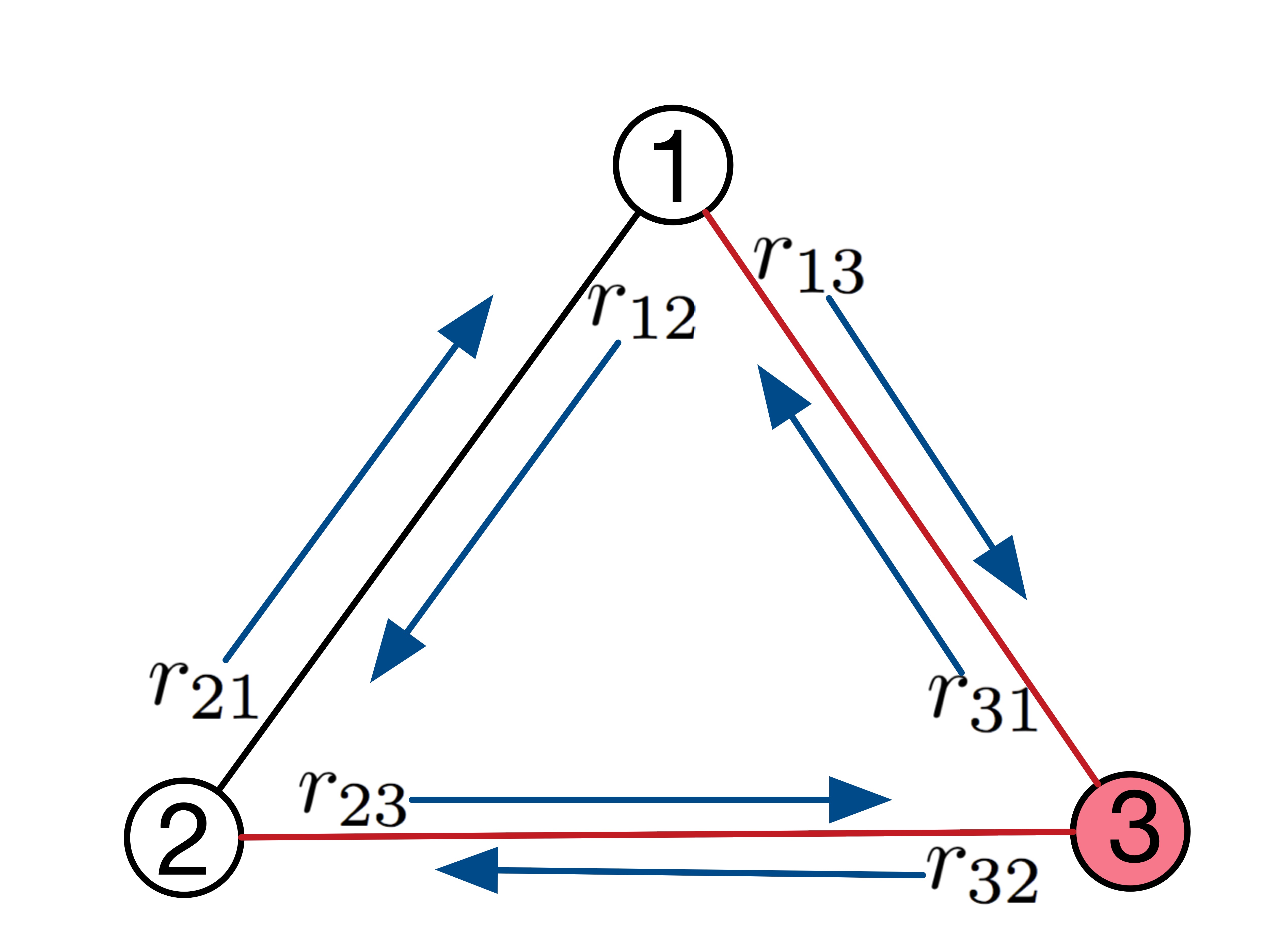}    
	\caption{\footnotesize{ Arrows (in blue) show the flow of information over an edge. }}
	\label{fig:illust}
\end{center}
\end{figure}

To demonstrate our proposed distributed average consensus protocol we consider a simple network of $3$ agents with $\mathcal{V} = \{1,\, 2, \, 3\}$ and $\mathcal{E} = \left\{ \{1,\,2\}, \, \{1,\,3\}, \, \{2,\,3\} \right\}$, as shown in Fig. \ref{fig:illust}.  

Let $h_1(x) = (x - x_1)^2$, $h_2(x) = (x - x_2)^2$ and $h_3(x) = (x - x_3)^2$, where $x \in \R$ is the public variable and $x_i \in \R, \, \forall i$ are the private constants held by agents.  

Let $\C = \{3\}$ and so, $\H = \{1, \, 2\}$. If the agents use a consensus-based gradient method~\cite{nedic2001distributed} for distributed optimization of the aggregate $\sum_{i}h_i(x)$ then the adversarial agent $3$ acquires knowledge of $x_1$ and $x_2$ (using the received optimal estimates of $1$ and $2$ at each time-step of the optimization algorithm). Now, to prevent this loss of privacy, the agents implement the proposed privacy protocol in the following manner.
 
In first phase, the agents execute the following steps
\begin{enumerate}
        \item 
        As shown in Fig. \ref{fig:illust}, all pairs of adjacent agents $i$ and $j$ exchange the respective values $r_{ij}$ and $r_{ji}$ (chosen independently and following a normal distribution, with mean $0$ and variance $\sigma = 1$, in $\mathbb{R}$) with each other. Consider a particular instance where  
        \begin{align*}
        	&r_{12} = 0.1, \, r_{21} = 0.5, \, r_{23} = 0.7, \, r_{32} = 0.4, \, r_{31} = 0.3 \\
        	&r_{13} = 0.8
        \end{align*}
        \item
            The agents compute their respective masks, 
            \begin{align*}
            	a_1 =  (r_{21} - r_{12}) + (r_{31} - r_{13}) = -0.1
            \end{align*}
            Similarly, $a_2 = -0.7$ and $a_3 = 0.8$. (One can verify that $(a_1 + a_2 + a_3) = 0$.)
        \item 
            The agents compute their respective effective costs,
            \begin{align*}
            	& \widetilde{h}_1(x) = (x-x_1)^2 - 0.1x = x^2 - (2x_1 + 0.1)x + x_1^2
            \end{align*}
            Similarly, $\widetilde{h}_2(x) = x^2 - (2x_2 + 0.7)x + x_2^2$ and $\widetilde{h}_3(x) = x^3 - (2x_3 - 0.8)x + x_3^2$.
\end{enumerate}

After the first phase, each agent uses a (non-private) distributed optimization algorithm $\Pi$ in the second phase to optimize $\sum_{i}\widetilde{h}_i$ (it can be easily to verified that $ \sum_{i} \widetilde{h}_i \equiv  \sum_{i} h_i$).

Here, as agent $3$ does not cut the honest agents $1$ and $2$, therefore agent can only determine $2x_1 + 2x_2$ with certainty and not the individual values of $x_1$ and $x_2$ (cf. Theorem \ref{thm:priv_2}). Specifically, as $\sigma = 1$ and $\underline{\mu}(\L_\H) = 2$
\begin{align*}
    &D_{KL}(\view_{\{3\}}(2x_1, 2x_2)||\view_{\{3\}}(2x'_1, 2x'_2)) \\ 
    &\leq \frac{1}{c}\{(x_1 - x'_1)^2 + (x_2 - x'_2)^2\}
\end{align*}
for any two $(x_1, x_2)$ and $(x'_1, x'_2)$ that satisfy 
\[x_1 + x_2 = x'_1 + x'_2\]

\section{Concluding Remarks}
\label{sec:dis}

In this paper, we have proposed a protocol for protecting privacy of the agents' costs, in peer-to-peer distributed optimization, against passive adversaries that corrupt some fraction of agents in the network. The proposed protocol preserves the privacy of the affine parts of the honest (not corrupted by the adversary) agents' costs if the corrupted agents do not constitute a vertex cut of the network. The only information that the adversary can get on the affine parts of the honest agents' costs is their aggregate. This implies that the proposed protocol can guarantee privacy of the affine parts of the honest agents' costs against a passive adversary that corrupts at most $t$ agents if the communication network has $(t+1)$-connectivity.

\bibliographystyle{IEEEtran}
\bibliography{references_optimization}

\end{document}